\documentclass{pasj01}

\pdfoutput=1 

\Received{$\langle$reception date$\rangle$}
\Accepted{$\langle$acception date$\rangle$}
\Published{$\langle$publication date$\rangle$}

\usepackage{graphicx}
\usepackage{natbib}
\usepackage{datetime}
\usepackage{url}
\usepackage{lscape}    
\usepackage{longtable}     

\usepackage{xcolor}
\definecolor{applegreen}{rgb}{0.55,0.71,0.0}
\definecolor{forestgreen}{rgb}{0.13,0.55,0.13}
\definecolor{pinegreen}{rgb}{0.0,0.47,0.44}
\definecolor{upforestgreen}{rgb}{0.0,0.27,0.13}
\definecolor{vividviolet}{rgb}{0.62,0.0,1.0}
\definecolor{lightcyan}{rgb}{0.717647, 0.933071, 0.996078} 
\definecolor{scarlet}{rgb}{1.0, 0.13, 0.0}

 \newcommand\ourowncheck{$\surd$}

\long\def\comment#1{}
\comment{
   Usage:
     \comment{ comments....}
}

\usepackage{bm}
\usepackage{ulem}
\DeclareRobustCommand{\erase}{\bgroup\markoverwith{\textcolor{red}{\rule[.5ex]{2pt}{0.6pt}}}\ULon}
 \newcommand{\add}[1]{\textcolor{black}{#1}}

 \def\noflightsource{60,000}
\def\noflightsourceTotal{240,000}

\begin{document}

\title{A deep analysis for New Horizons' KBO search images}

\author{Fumi Yoshida${}^{1,2,\ast}$}\email{fumi-yoshida@med.uoeh-u.ac.jp}
\author{Toshifumi Yanagisawa${}^{3}$}
\author{Takashi Ito${}^{4,2,8}$}
\author{Hirohisa Kurosaki${}^{3}$}
\author{Makoto Yoshikawa${}^{5}$}
\author{Kohki Kamiya${}^{3}$}
\author{Ji-an Jiang${}^{6,7}$}
\author{Alan Stern${}^{9}$}
\author{Wesley C. Fraser${}^{10,11}$}
\author{Susan D. Benecchi${}^{12}$}
\author{Anne J. Verbiscer${}^{13,9}$}

\altaffiltext{1}{University of Occupational and Environmental Health, Japan, 1--1 Iseigaoka, Yahata, Kitakyusyu 807--8555, Japan}
\altaffiltext{2}{Planetary Exploration Research Center, Chiba Institute of Technology, 2--17--1 Tsudanuma, Narashino, 275--0016, Chiba, Japan}
\altaffiltext{3}{Chofu headquarters, Japan Aerospace Exploration Agency, Jindaiji Higashimachi 7--44--1, Chofu, Tokyo 182--0012, Japan}
\altaffiltext{4}{Center for Computational Astrophysics, National Astronomical Observatory of Japan, Osawa 2--21--1, Mitaka, Tokyo 181--8588, Japan}
\altaffiltext{5}{ISAS, Japan Aerospace Exploration Agency, Yoshinodai 3--1--1, Sagamihara, Kanagawa 252--0222, Japan}
\altaffiltext{6}{Department of Astronomy, University of Science and Technology of China, Hefei 230026, China}
\altaffiltext{7}{Division of Science, National Astronomical Observatory of Japan, Osawa 2--21--1, Mitaka, Tokyo 181--8588, Japan}
\altaffiltext{8}{College of Science and Engineering, Chubu University, 1200 Matsumoto--cho, Kasugai, 487--8501, Aichi, Japan}
\altaffiltext{9}{Southwest Research Institute, 1050 Walnut Street, Boulder, CO 80302, USA}
\altaffiltext{10}{National Research Council of Canada, Herzberg Astronomy and Astrophysics Research Centre, 5071 W. Saanich Rd., Victoria, BC, V9E 2E7, Canada}
\altaffiltext{11}{Department of Physics and Astronomy, University of Victoria, Elliott Building, 3800 Finnerty Road, Victoria, BC, V8P 5C2, Canada}
\altaffiltext{12}{Planetary Science Institute, 1700 East Fort Lowell, Suite 106, Tucson, AZ 85719, USA}
\altaffiltext{13}{Department of Astronomy, University of Virginia, P.O. Box 400325, Charlottesville, VA 22904--4325, USA}


\maketitle

\begin{abstract}
Observation datasets acquired by the Hyper Suprime-Cam (HSC) on the Subaru Telescope for NASA's New Horizons mission target search were analyzed through a method devised by JAXA.
The method makes use of Field Programmable Gate arrays and was originally used to detect fast-moving objects such as space debris or near-Earth asteroids.
Here we present an application of the method to detect slow-moving Kuiper Belt Objects (KBOs) in the New Horizons target search observations.
A cadence that takes continuous images of one HSC field of view for half a night fits the method well.
The observations for the New Horizons Kuiper Belt Extended Mission (NH/KEM) using HSC began in May 2020, and are ongoing.
Here we show our result of the analysis of the dataset acquired from May 2020 through June 2021 that have already passed the proprietary period and are open to the public.
We detected 84 KBO candidates in the June 2020 and June 2021 datasets, when the observation field was close to opposition.
\end{abstract}


\comment{
\begin{center}
{\Huge \textcolor{magenta}{\textbf{Last updated: \\
{\currenttime} {\today}}}}
\end{center}
\tableofcontents
}


\section{Introduction} \label{sec:intro}

Trans-Neptunian Objects (TNOs) are a diverse group of small bodies that exist beyond Neptune's orbit.
The first discovered TNO was Pluto.
Since the discovery of other objects in this region \citep{jewitt1993}, people began to realize the existence of a swarm of small bodies and dwarf planets in a belt-like zone beyond Neptune's orbit.
The objects in this belt with semimajor axis $\sim$30--50 au are called the Kuiper Belt Objects (KBOs).
KBOs are roughly divided into four distinct dynamical groups: resonant KBOs, classical KBOs, scattered (or scattering) disk objects, and detached objects.
The resonant KBOs are trapped in various mean motion resonances with Neptune.
The classical KBOs are not in strong mean motion resonance with Neptune, and their orbits are overall nearly circular.
KBOs with perihelion distances larger than 30 au and with large eccentricities are the scattered disk objects (SDOs).
The detached objects are those that have larger perihelion distances than the SDOs and are less affected by Neptune's gravity.
For additional details on the classification and nomenclature of KBOs, consult \citet{gladman2008}.

To unravel the dynamical structure of the solar system beyond Neptune and to understand the physical processes that formed this structure, it is crucial to discover more objects and to better characterize their physical properties.
For this purpose, both ground-based observations and spacecraft exploration are necessary; they are complementary to each other.
Only a few spacecraft have traversed the Kuiper Belt; however, NASA's New Horizons is the only spacecraft that has directly observed KBOs at close range and is currently operational.

\add{%
New Horizons is a mission designed for flyby exploration of Pluto and other KBOs \citep{stern2003}.
The spacecraft was launched in January 2006, reached the Pluto system in July 2015, and then made a close flyby of one of the classical KBOs, (486958) Arrokoth, in January 2019 \citep{stern2015,stern2019}.
After the Pluto flyby, the mission entered extended phases during which the spacecraft observed many KBOs as point sources \citep{porter2016,verbiscer2019,verbiscer2022} while the team searched for a flyby target to follow Arrokoth \citep{stern2018}.} 
From 2004--2014, the science team used the Subaru Telescope's Suprime-Cam (SC: the predecessor of HSC; \citet{miyazaki2002}) to search for the initial post-Pluto flyby candidate \citep{buie2024}.
In May 2020, the science team resumed the ground-based survey observations 
to search for the next flyby target as well as for KBOs observable from the spacecraft.
One of the major facilities used for this purpose is the Subaru Telescope (\url{https://subarutelescope.org/en/}), and it was at this time when members from Japan participated in the science team and began playing roles.
One of the main objectives of the science team, in particular with the participation of Japan, is to exploit the Subaru Telescope and its wide field camera, Hyper Suprime-Cam (hereafter HSC), to discover more KBOs for the spacecraft to observe, either as point sources or at high resolution during a close flyby.

Currently, the spacecraft is beyond the so-called Kuiper cliff (the outer edge of the Kuiper Belt) which has been proposed to be located at $\sim 50$ au from the Sun \citep[e.g.][]{chiang1999,delafuentemarcos2024}.
The ground-based observations that started in May 2020 have not yet found any objects that the New Horizons spacecraft can reach for a close flyby.
However, the data obtained from the ground-based observations to date is by no means useless because the images contain many KBOs with substantial scientific importance.
The purpose of the present paper lies here {\textemdash} we describe our effort to exploit the ground-based observational data and present our initial detections as well as our methods. 
In general, detecting KBOs from ground-based observations is a formidable task because of their faintness and the complicating effects of background star confusion.
Different analysis methods can yield somewhat different outcomes even if they are based on the same observational data.
In this sense, our present analysis, which is independent of but complementary to the effort of the main New Horizons team in North America, is significant because it provides an independent, objective view of the dataset and a chance to recover objects missed through other search efforts.

In Section \ref{sec:observation} we give specific information on the ground-based observations that resulted in the dataset we use for this study.
In Section \ref{sec:method} we describe in detail how we detect moving objects in the dataset.
Section \ref{sec:link} is devoted to describing how we connect the orbits of a moving object over multiple observation nights.
Section \ref{sec:discussion} provides a discussion and summary of this study.

 \section{Observation dataset and reduction} \label{sec:observation}

All of the observation datasets whose results are presented here were obtained using HSC on the Subaru Telescope.
We select two fields of view (hereafter referred to as F1 and F2) that include New Horizon's flight path.
At each observation opportunity, one of the fields is imaged continuously for half a night principally with a constant exposure time of 90 seconds.
On the next night, the other field is imaged in the same manner.
This procedure is followed throughout the entire observation period.
After an interval of $\sim 30$ days, whenever possible, another observation is repeated in the same way to extend the observation arc.
This refines and enhances the accuracy of orbit determination of the detected objects.

In Table \ref{tbl:obslog}, we summarize the details of the observational datasets obtained for the New Horizons mission since May 2020; these observations are publicly available as of September 2023.
Among them, we analyzed the 14 sets (marked with $\surd$ in the leftmost column of the table).
This table lists, from the left,
which dataset we used ($\surd$),
observation date (UT),
reduction ID (red. id),
start time of the first and the last exposures,
position of the field of view (RA2000, DEC2000),
field id (either F1 or F2),
filter,
exposure time (s),
number of acquired images, and
associated information on the night condition.
Note that the observation datasets listed in Table \ref{tbl:obslog} are all beyond the proprietary limits of the original observers and, are publicly available the SMOKA public archive site.
See \citet{baba2002} for more details on SMOKA or visit their website, \url{https://smoka.nao.ac.jp/}.
\add{Note also that Table \ref{tbl:obslog} is also available in the Excel format as Supplementary Data.}

HSC is a huge mosaic CCD camera mounted on the prime focus of the Subaru Telescope.
HSC has 116 CCDs with a pixel size of 15 $\mu$m and pixel scale of $0.168''$.
Among the 116 CCDs, 104 of them are used for scientific observation, 4 are for the auto guider, and 8 are for focus monitoring
\add{%
(note that one of the 104 chips for scientific observation is not functional as of 2024 April).}
HSC is designed to conduct large-scale sky surveys with significant depth and high precision with a large field of view of $\sim 1^\circ.5$ diameter.
The image quality across the field of view has a median seeing size of $\sim 0.6''$ in the $i$-band \citep{aihara2018a,aihara2018b,aihara2019,aihara2022}.
For technical details of HSC, see
\citet{miyazaki2018} for its system design,
\citet{komiyama2018} for its camera dewar design,
\citet{kawanomoto2018} for its filter system, and
\citet{furusawa2018} for its on-site quality-assurance system.
A dedicated website is also available: \url{https://www.subarutelescope.org/Observing/Instruments/HSC/}.

Most of the observations were acquired on half nights; a few were acquired on quarter nights. We used the $r2$ filter except on 2020 August 26 when we used the $z$-band filter due to a request from the observatory.
The sky condition described in Table \ref{tbl:obslog} was transcribed from the observation night log that the Subaru Telescope operators reported. This is included because it helps us to evaluate the quality of the acquired data.

For the acquired data, standard image reduction procedures including bias, dark, flat, and fringe corrections, as well as astrometric and photometric calibrations against the Pan-STARRS1 (PS1) $3\pi$ catalog \citep{tonry2012,magnier2013} were carried out using the HSC pipeline, a version of the Large Synoptic Survey Telescope (LSST) stack \citep{ivezic2019,axelrod2010}.
We refer the reader to \citet{aihara2018a} and \citet{jiang2020} for further details on the HSC data reduction.

This study only used 14 of the 22 datasets listed in Table \ref{tbl:obslog} for 
two major reasons.
First, although $\sim$100 images of the same field were continuously taken each night, we did not find a sequence of 32 images with a constant time interval (e.g. the set with Reduction ID $03068$) which, as mentioned in Section \ref{sec:method}, our detection method requires; therefore, we cannot apply it to these irregular image sequences. Currently, dealing with this issue is beyond the scope of this work.
There are several possible causes for the temporal inconsistency in the image sequences, including refocusing the telescope according to changing weather conditions, or interruptions to the observation sequence due to unexpected instrumental trouble during an exposure, resulting in irregular time intervals between images.

Moreover, even if a sequence of 32 images with a constant time interval is available, if more than half of the images were taken under bad sky conditions with poor seeing, our detection method fails. 
This can happen at any time.
However, unlike the first issue (i.e. the inconsistencies due to considerations of telescope operations), we believe we can address the second issue in the near future by considering seeing differences in individual image subtractions.
We expect this will come at the cost of a shallower limiting magnitude, but more of the dataset will be usable.
We discuss more about this modification in Section \ref{sec:discussion}.

\onecolumn
{\footnotesize
\begin{landscape}
\begin{center}
\begin{longtable}[c]{cccccccclccl}
\caption{
Description of the observational datasets obtained for the New Horizons mission conducted at the Subaru Telescope with HSC from May 2020 through June 2021.
Among the 22 datasets, we analyzed 14 sets identified by $\surd$ in the leftmost column.
\add{This table is also available in the Excel format as Supplementary Data.}
}
\\
\hline 
 &
DATEOBS & red. & start time & start time & RA2000 & DEC2000 & field & filter & exp & {\#} & \\
& (UT)          &    id & of the first             &  of the last   &   (deg)      &     (deg)   &    id       &              &    time  & of & \multicolumn{1}{c}{night condition} \\
&                  &                   & exposure         &  exposure              &                  &                &            &               &  (s)         & images&\\
\hline
\hline 
 &
2020/05/26	& 03068 &	10:37:11.962	&	13:19:52.429	&	287.376 	&	$-20.227$ 	&	F1	&	$r2$	&	90	&	80	&\small{(Sky) cloudy, (Seeing) 0.7--1.2$\arcsec$} \\
 & &&&&&&&&&& \small{(Temp) 5.7--7.0\degree C, (Wind) 1.0--7.4 m/s }\\
 & &&&&&&&&&&\small{(Humid) 0.5--19.9\%} \\
\ourowncheck &
2020/05/28	& 03070 &	10:42:33.966	&	13:22:08.568	&	288.740 	&	$-20.584$ 	&	F2	&	$r2$	&	90	&	78	&\small{(Sky) cloudy, (Seeing) 0.7--0.9$\arcsec$ }\\
 & &&&&&&&&&&\small{(Temp) 5.4--6.3\degree C, (Wind) 1.9--13.2 m/s}  \\
 & &&&&&&&&&& \small{(Humid) 3.7--14.0\%}\\
\ourowncheck &
2020/05/29	& 03071&	10:42:54.138	&	14:50:41.399	&	287.338 	&	$-20.127$ 	&	F1	&	$r2$      &	90	&	124	&\small{(Sky) clear, (Seeing) 0.5--0.9$\arcsec$ }\\
 & &&&&&&&&&& \small{(Temp) 4.8--5.4\degree C, (Wind) 1.9--15.4 m/s }\\
 & &&&&&&&&&&\small{(Humid) 6.9--31.5\%}\\
\ourowncheck &
2020/05/30	& 03072 &	10:31:55.756	&	14:50:04.288	&	288.714 	&	$-20.380$ 	&	F2	&	$r2$	&	90	&	128	&\small{(Sky) clear, (Seeing) 0.7--1.2$\arcsec$}\\
 & &&&&&&&&&&\small{(Temp) 3.6--5.9\degree C, (Wind) 2.5--15.4 m/s}\\
 & &&&&&&&&&&\small{(Humid) 17.5--28.5\%}\\
\ourowncheck &
2020/05/31	& 03073 &	10:34:52.068	&	14:49:22.538	&	287.312 	&	$-20.129$ 	&	F1	&	$r2$	&	90	&	128	&\small{(Sky) clear, (Seeing) 0.6--0.9$\arcsec$ }\\
 & &&&&&&&&&&\small{(Temp) 3.4--5.5\degree C, (Wind) 2.7--12.5 m/s} \\
 & &&&&&&&&&&\small{(Humid) 7.4--17.2\%}\\
 &
2020/06/01	& 03074 &	10:22:19.511	&	14:37:33.060	&	288.688 	&	$-20.382$ 	&	F2	&	$r2$	&	90	&	62	&\small{(Sky) clear, (Seeing) 0.55--0.9$\arcsec$ }\\
 & &&&&&&&&&&\small{(Temp) 3.0--6.2\degree C, (Wind) 2.5--17.8 m/s}\\
 & &&&&&&&&&&\small{(Humid) 29.2--85.6\%}\\
 &
2020/06/19	& 03092 &	10:31:59.346	&	14:56:30.976	&	287.145 	&	$-20.042$ 	&	F1	&	$r2$	&	90	&	131	&\small{(Sky) cloudy, (Seeing) 0.9--1.6$\arcsec$} \\
 & &&&&&&&&&& \small{(Temp) 2.0--3.6\degree C, (Wind) 2.2--19.2 m/s}\\
 & &&&&&&&&&& \small{(Humid) 5.3--9.7\%}\\
\ourowncheck &
2020/06/20	& 03093 &	10:55:39.175	&	14:55:24.839	&	288.662 	&	$-20.049$ 	&	F2	&	$r2$	&	90	&	118	&\small{(Sky) high clouds, (Seeing) 0.7--1.1$\arcsec$ }\\
 & &&&&&&&&&& \small{(Temp) 6.9--7.6\degree C, (Wind) 1.9--7.7 m/s}\\
 & &&&&&&&&&& \small{(Humid) 2.2--24.4\%}\\
\ourowncheck &
2020/06/21	& 03094 &	10:52:52.432	&	14:41:24.850	&	287.112 	&	$-20.059$ 	&	F1	&	$r2$	&	90	&	114	&\small{(Sky) clear, (Seeing) 0.7--1.0$\arcsec$ }\\
 & &&&&&&&&&&\small{(Temp) 5.2--6.8\degree C, (Wind) 6.8--21.6m/s} \\
 & &&&&&&&&&&\small{(Humid) 2.6--4.8\%}\\
\ourowncheck &
2020/06/22	& 03095 &	10:46:30.525	&	14:10:33.646	&	288.629 	&	$-20.054$ 	&	F2	&	$r2$	&	90	&	102	&\small{(Sky) thin cirrus, (Seeing) 0.56--0.7$\arcsec$ }\\
 & &&&&&&&&&& \small{(Temp) 6.2--7.4\degree C, (Wind) 5.5--13.7m/s} \\
 & &&&&&&&&&& \small{(Humid) 1.5--3.3\%}\\
\ourowncheck &
2020/06/24	& 03097 &	10:26:47.483	&	14:47:51.861	&	287.062 	&	$-20.064$ 	&	F1	&	$r2$	&	90	&	130	&\small{(Sky) clear, (Seeing) 0.66--0.8$\arcsec$ }\\
 & &&&&&&&&&& \small{(Temp) 3.3--6.2\degree C, (Wind) 1.9--13.7 m/s}\\
 & &&&&&&&&&& \small{(Humid) 3.6--8.0\%} \\
\ourowncheck &
2020/06/25	& 03098 &	10:44:42.581	&	14:44:40.567	&	288.578 	&	$-20.060$ 	&	F2	&	$r2$	&	90	&	120	&\small{(Sky) clear, (Seeing) 0.65--2.0$\arcsec$ }\\
 & &&&&&&&&&& \small{(Temp) 3.1--6.4\degree C, (Wind) 2.0--12.4 m/s} \\
 & &&&&&&&&&&\small{(Humid) 2.3--12.9\%}\\
\ourowncheck &
2020/08/12	& 03146 &	06:23:26.824	&	10:17:44.506	&	286.251 	&	$-20.165$ 	&	F1	&	$r2$	&	90	&	117	&\small{(Sky) clear, (Seeing) 0.7--1.1$\arcsec$} \\
 & &&&&&&&&&& \small{(Temp) 1.4--4.6\degree C, (Wind) 1.9--6.6 m/s}\\
 & &&&&&&&&&& \small{(Humid) 13.5--28.6\%}\\
\ourowncheck &
2020/08/13	& 03147 &	05:49:38.806	&	10:08:23.645	&	287.676 	&	$-20.123$ 	&	F2	&	$r2$	&	90	&	129	&\small{(Sky) clear, (Seeing) 0.7--1.1$\arcsec$ }\\
 & &&&&&&&&&& \small{(Temp) 1.8--4.7\degree C, (Wind) 1.9--13.6 m/s}\\
 & &&&&&&&&&& \small{(Humid) 27.8--58.8\%}\\
\ourowncheck &
2020/08/14	& 03148 &	05:47:57.363	&	10:05:59.358	&	286.223 	&	$-20.169$ 	&	F1	&	$r2$	&	90	&	128	&\small{(Sky) clear, (Seeing) 0.5--1.2$\arcsec$} \\
 & &&&&&&&&&&  \small{(Temp) 2.9--3.2\degree C, (Wind) 2.2--6.7 m/s}\\
 & &&&&&&&&&&  \small{(Humid) 39.9--43.9\%}\\
 &
2020/08/15	& 03149 &	05:49:37.187	&	10:05:39.643	&	287.743 	&	$-20.164$ 	&	F2	&	$r2$	&	90	&	122	&\small{(Sky) clear, (Seeing) 0.6--1.3$\arcsec$ }\\
 & &&&&&&&&&& \small{(Temp) 3.2--4.6\degree C, (Wind) 1.9--9.0 m/s }\\
 & &&&&&&&&&& \small{(Humid) 17.0--35.7\%}\\
 &
2020/08/25	& 03159 &	05:48:22.166	&	10:01:32.822	&	286.085 	&	$-20.177$ 	&	F1	&	$r2$	&	90	&	124	&\small{(Sky) clear, (Seeing) 1.0--1.5$\arcsec$ }\\
 & &&&&&&&&&& \small{(Temp) 1.1--2.7\degree C, (Wind) 5.1--18.0 m/s}\\
 & &&&&&&&&&& \small{(Humid) 5.3--32.1\%}\\
 &
2020/08/26	& 03160 &	05:41:09.488	&	10:13:26.988	&	286.073 	&	$-20.179$ 	&	F1	&	$z$	&	90	&	134	&\small{(Sky) clear, (Seeing) 1.4--3.5$\arcsec$} \\
 & &&&&&&&&&& \small{(Temp) 3.0--3.9\degree C, (Wind) 1.9--16.9 m/s} \\
 & &&&&&&&&&& \small{(Humid) 12.2--17.0\%}\\
 &
2020/10/10	& 03205 &	05:04:30.794	&	06:45:04.382	&	287.226 	&	$-20.105$ 	&	F2	&	$r2$	&	90	&	51	&\small{(Sky) cirrus, (Seeing) 0.4--1.2$\arcsec$} \\
 & &&&&&&&&&&  \small{(Temp) 1.6--4.0\degree C, (Wind) 2.0--15.7 m/s} \\
 & &&&&&&&&&& \small{(Humid) 7.0--76.6\%}\\
 &
2020/10/12	& 03207 &	04:53:27.086	&	06:55:31.830	&	287.227 	&	$-20.105$ 	&	F2	&	$r2$	&	90	&	61	&\small{(Sky) clear to overcast, (Seeing) 0.6--1.1$\arcsec$} \\
 & &&&&&&&&&& \small{(Temp) 6.0--6.8\degree C, (Wind) 3.8--12.4 m/s} \\
 & &&&&&&&&&& \small{(Humid) 5.7--7.8\%}\\
\ourowncheck &
2021/06/09	& 03447 &	10:35:44.164	&	14:43:46.423	&	288.550 	&	$-21.100$ 	&	F2	&	r2	&	90	&	122	&\small{(Sky) clear, (Seeing) 0.58--1.15$\arcsec$} \\
 & &&&&&&&&&& \small{(Temp) 1.2--3.7\degree C, (Wind) 6.9--16.9m/s} \\
 & &&&&&&&&&& \small{(Humid) 9.0--21.5\%}\\
\ourowncheck &
2021/06/17	& 03455 &	10:27:09.891	&	13:53:45.662	&	288.450 	&	$-21.100$ 	&	F2	&	r2	&	90	&	93	&\small{(Sky) clear to overcast, (Seeing) 0.28--2.91$\arcsec$}\\
 & &&&&&&&&&& \small{(Temp) $-2.7$--$-0.6$\degree C, (Wind) 2.0--10.8 m/s}\\
 & &&&&&&&&&& \small{(Humid) 51.8--90.5\%}\\
\hline 
\label{tbl:obslog}
\end{longtable}
\end{center}
\end{landscape}
} 
\twocolumn

\section{Detection of moving objects} \label{sec:method}

The datasets described in Section \ref{sec:observation} are well-suited for the moving object detection method that our group has developed \citep[e.g.][]{yanagisawa2005,yanagisawa2021},
the JAXA (Japan Aerospace Exploration Agency) detection method.
The JAXA detection method superimposes images based on an assumed motion of moving objects possibly embedded in the images.
When a moving object is embedded in a sequence of images, and if its motion vector is consistent with the assumed trajectory, we can make the object appear much brighter and more visible (i.e. increasing the signal to noise ratio) by superimposing images together with median filtering.

This method has mainly been used for detecting space debris around the Earth and faint near-Earth objects that are not visible in a single image \citep[e.g.][]{yanagisawa2012b,yanagisawa2015}.
\add{%
Detection of such objects is difficult not only because many of these are faint, but also because their on-sky velocity is sometimes very large \citep[e.g.][]{ito2010}.
On the other hand, in this study, we attempt to apply this detection method to find KBOs whose motion is much slower than that of space debris or near-Earth objects.}
The efficiency of this method has recently been improved by implementing binarization of images and hard-coding the algorithm on an FPGA board.
This method employs 32 images of the same field taken at a constant, equal time interval.
The series of observational datasets acquired for New Horizons described in Section \ref{sec:observation} largely fulfills the condition.
\add{Note that in this study, we define binarization as an image processing technique that converts a grayscale image into an image containing only two pixel values.
}

\begin{figure}[ht!]\centering
  \includegraphics[width=\textwidth]{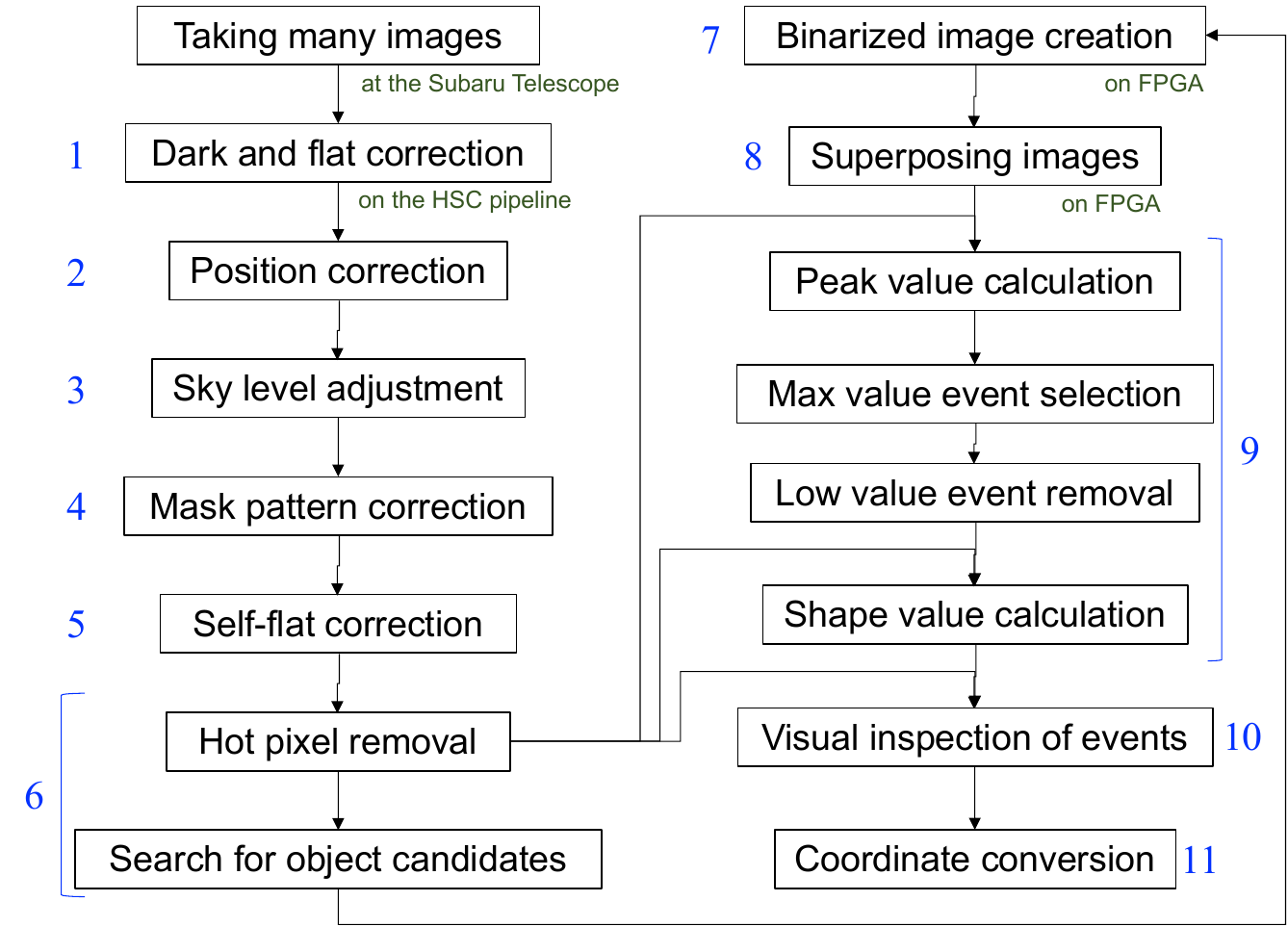}
\caption{%
Flowchart of the object detection procedures.
The starting box is ``Taking many images'' at the top left.
The numbers in blue denote the task number described in the main text.
\label{fig:yanagi-flowchart}}
\end{figure}

\begin{figure}[ht!]\centering
  \includegraphics[width=0.96\textwidth]{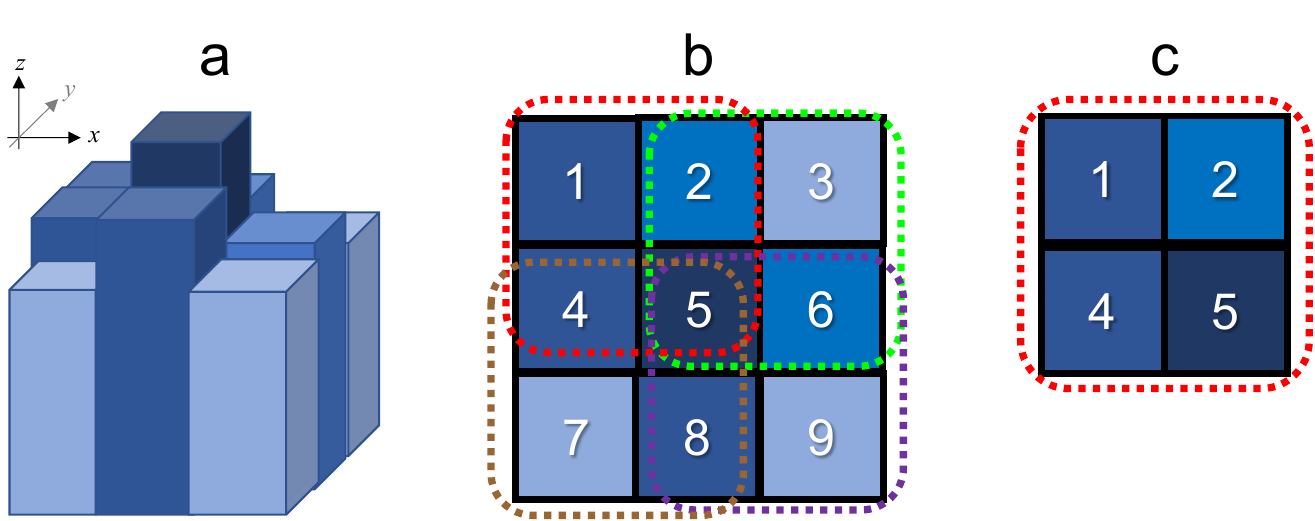}
\caption{%
\add{%
Schematic illustration of how we binarize a $2 \times 2$ pixel set
in the tasks 6 and 7.
Note that this illustration is just about one light source (i.e. one moving object candidate) on a single image.
Panel \textsf{a} shows the pixel values of the nine pixels around the light source and their positions in the three-dimensional space, $(x,y,z)$.
$(x, y)$ are the coordinates of the pixels on the image, and $z$ contains their pixel values.
Panel \textsf{b} shows the top view of the panel \textsf{a} with the numbers on each pixel from 1 to 9.
We calculate the sum values of each of the $2 \times 2$ pixel sets (shown in the dotted lines; 4 sets in total).
Panel \textsf{c} shows the $2 \times 2$ pixel set whose sum of the pixel values are the largest, and we set their pixel values to 1.
We set the value of all other pixels $(3,6,7,8,9)$ to 0.}
\label{fig:yanagi-select_4pix}}
\end{figure}

\begin{figure}[!ht]\centering
  \includegraphics[width=\textwidth]{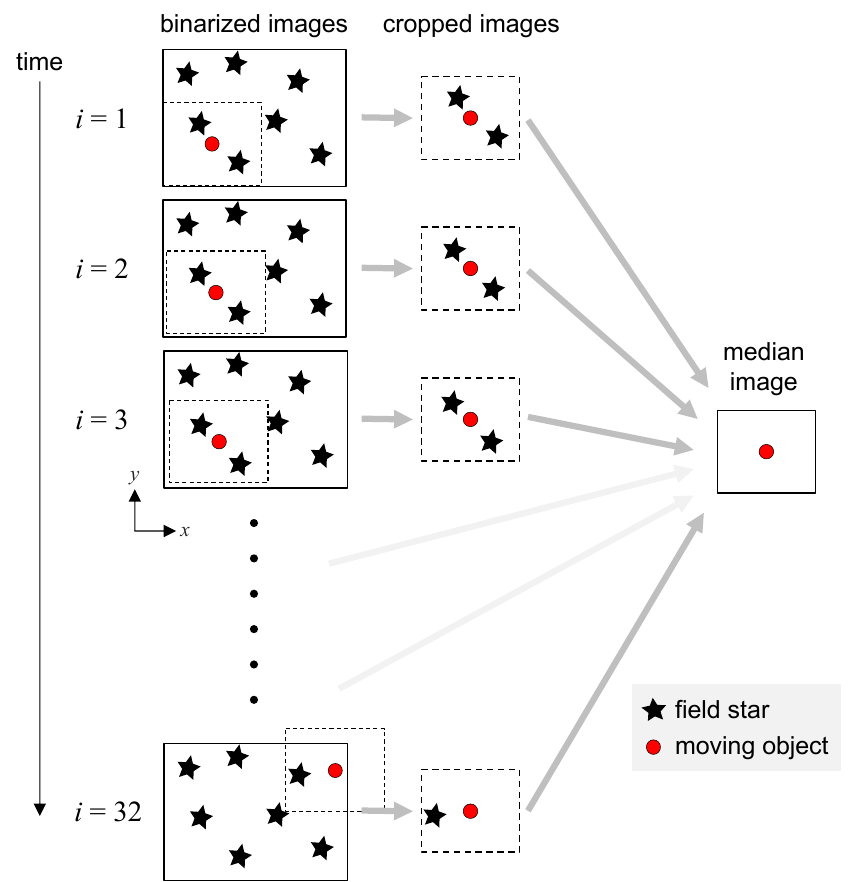}
\caption{%
Schematic illustration of task 8 (cropping and stacking images).
\add{%
Time proceeds from the top to the bottom of the figure.
The index $i$ (from 1 to 32) denotes the number among the 32 images we used for the detection procedure.
The coordinate axis $({x}, {y})$ at the middle left of the figure schematically represent the coordinates on the images.
}
\label{fig:yanagi-detect_schem}}
\end{figure}

In what follows we briefly describe how this method works.
We summarized the flow of the tasks in Figure \ref{fig:yanagi-flowchart} with the supporting illustration in Figures \ref{fig:yanagi-select_4pix} and \ref{fig:yanagi-detect_schem}.
From the series of images acquired on a single night, we select a set of 32 images obtained with a constant time interval.
Note that we select the 32 images out of the entire $\sim 100$ images per night so that the time length covered by the selected images is as long as possible during the observation night while maintaining a fixed time interval.
Then we carry out the following tasks (1--11).

\begin{enumerate}
\item The HSC pipeline \citep{bosch2018} carries out bias subtraction, flat correction, and sky subtraction.
\item Since the pointing of the telescope varies slightly (or is off) from image to image, we align the images to each other.
\item We create a median image of the 32 aligned images.
We subtract the median image from each of the 32 images.
\item We mask each image with the mask patterns created from the median image.
We set the masking threshold at 10 times higher than the average noise values.
\item
We create the self-flat images through the $\sigma$-clipping procedure (\url{https://www.gnu.org/software/gnuastro/manual/html_node/Sigma-clipping.html}) as well as by smoothing each image whose mask pattern has been corrected.
We then divide each image by its self-flat image.
This task removes the sky level pattern that varies with atmospheric turbulence.
\item
\add{%
We search for moving object candidates in the masked images.
For this task and the next, we introduce a parameter named the shape parameter (hereafter denoted as $P_\mathrm{shape}$).
In a $3 \times 3$ pixel set that contains a light source, $P_\mathrm{shape}$ is defined as the value obtained through an operation of dividing the sum of the values of the $3 \times 3 = 9$
pixels around the peak of the light source (including the peak itself) by the peak pixel value (see Figure \ref{fig:yanagi-select_4pix} and the explanation about it below).}
When $P_\mathrm{shape} \sim 1$, only the central pixel has an exceptionally high value (the so-called ``hot pixel'').
This type of light source is probably caused by dead pixels, noise, or cosmic rays.
When $P_\mathrm{shape} \gg 1$, we regard it as an object candidate at the position of an extended light source (i.e. a light source with the PSF having a finite extension).

\add{%
Here is the specific procedure to search for moving object candidates.
There are a large number of light sources in the images.
Among them, we extract 
about {\noflightsource} light sources with the shape parameters greater than 2 (i.e. $P_\mathrm{shape} > 2$).
Figure \ref{fig:yanagi-select_4pix}\textsf{a} is a schematic illustration of one of these light sources centered at a $3 \times 3$ pixel set.
Let us assume that these nine pixels have the three-dimensional coordinates, $(x,y,z)$,
where $(x, y)$ are the coordinates of each pixel in the image, and $z$ contains their pixel values.
Figure \ref{fig:yanagi-select_4pix}\textsf{a} illustrates a $3 \times 3$ pixel set on a single image that has $P_\mathrm{shape} > 2$ (i.e. a moving object candidate).
There are many other light sources like this, and in this way we collect $\sim${\noflightsource} of $3 \times 3$ pixel sets in each of the 32 image sets.
The number {\noflightsource} is empirically, but carefully, selected to maximize the efficiency of the JAXA detection method \citep{yanagisawa2021}.
If the number is too small, we would detect only bright moving objects.
If the number is too large, we would detect many false positives.
Then we sort the $\sim${\noflightsource} light sources by their central pixel value (i.e. the value of the pixel no.~5 in Figure \ref{fig:yanagi-select_4pix}\textsf{b}), and record their $(x,y)$ coordinates.
Most of the light sources are noise that random pattern of the sky background creates, but there may be true moving objects embedded among them.
}
\item 
\add{
Using the $\sim${\noflightsource} moving object candidates selected in the previous task, we create binarized images for each of the 32 image sets.
Specifically, for each of the $\sim${\noflightsource} candidates, we calculate the sum value of each $2 \times 2$ pixel set that contains the pixel of the object candidate.
As shown in Figure \ref{fig:yanagi-select_4pix}\textsf{b}, there are four $2 \times 2$ pixel sets for each of the object candidates.
Then we locate which of the $2 \times 2$ pixel set has the largest sum value, and set the values of its $2 \times 2 = 4$ pixels to 1.
As a result, the above procedure sets the pixel value of all the $2 \times 2$ pixel sets of the $\sim${\noflightsource} moving object candidates to 1.
In other words, about $2 \times 2 \times \noflightsource = \noflightsourceTotal$ pixels have the value of 1 in a single image.
This completes the binarization of a single image.
All subsequent tasks use a set of 32 images binarized in this way to detect moving objects.
}

Figure \ref{fig:yanagi-select_4pix} is a schematic illustration of how we binarize a $2 \times 2$ pixel set.
We calculate the sum of the pixel values in the four $2 \times 2$ pixel sets:
$\left( {1\; 2} \atop {4\; 5}\right)$,
$\left( {2\; 3} \atop {5\; 6}\right)$,
$\left( {4\; 5} \atop {7\; 8}\right)$, and
$\left( {5\; 6} \atop {8\; 9}\right)$.
If the sum of the pixel values of the $2 \times 2$ pixel set $\left( {1\; 2} \atop {4\; 5}\right)$ is the highest among them, we set their pixel values to 1.
We repeat this operation for all the {$\sim$\noflightsource} candidates on an FPGA board.
\item
Assuming a motion vector of the moving object population we want to detect (such as KBOs, main belt asteroids, near-Earth objects, and so on), we superimpose the binarized images with many shift patterns with $dx$ and $dy$ where ($x, y)$ are the coordinates on the images.
Both $dx$ and $dy$ range from $-256, -255, \ldots, +256$, and in total we have $(256+1+256)^2 = 263,169$ different shifts.
This operation is done in parallel on an FPGA board.

\add{%
A slightly more specific description as to what is done in this task is illustrated in Figure \ref{fig:yanagi-detect_schem}.
From each of the 32 images (from $i=1$ to 32 in Figure \ref{fig:yanagi-detect_schem}), we crop a region of a certain size to fit the assumed position and velocity of a moving object and produce an intermediate image as shown in the middle panels.
We stack the 32 intermediate images and create their median image.
If the 32 intermediate images were shifted and cropped along the exact position and velocity of a moving object,  the position of the moving object in the each intermediate image remains at the same place of the cropped images.
Therefore, if we stack all 32 intermediate images and create their median image, only the moving object will remain in the median image (rightmost cloumn of Figure \ref{fig:yanagi-detect_schem}).
}  

\item
In the previous task, some moving objects are detected more than once with different (but similar) shift patterns, in particular when the objects are bright.
We think this is caused by 
bright objects tend to have extended shapes on the image.
The peak pixel of a bright object is often surrounded by the pixels whose values are also large.
This results in ambiguity in determining the peak location of a bright object.
As a result, the motion vector of a bright object can be estimated in more than one shift pattern.

In such cases, we use the image from which we removed any hot pixels in task 6 (referred to as the hot-pixel-removed image) to determine the true detection event.
More specifically, we examine the peak value in the median image created using the hot-pixel-removed images for each event and consider the event with the largest peak value to be the true detection.
We interpret the shift value that causes the true detection event as the amount of the apparent motion of the detected (real) object.
\item
By assuming the velocity of a moving object, and by shifting the image in many directions, the superimposed images generated from task 8 allow us to find a faint moving object that could not be seen on a single image.
The superimposed images are checked by human eyes to determine whether they contain real objects or not (``human inspection'' or ``vetting'').
We show a pair of examples of the moving object detection on actual images in Figure \ref{fig:yanagi-visual_ex1} (a clean example) and in Figure \ref{fig:yanagi-visual_ex2}
\add{(a false detection example caused by a noise spike; the noise in each image is superimposed on each other by chance).}
\add{Clean detections and false detections caused by noise spikes are generally distinguished as follows.}
When an actual moving object is detected as in Figure \ref{fig:yanagi-visual_ex1}, its light source is spread out on the image and multiple pixels have large values.
On the other hand, in Figure \ref{fig:yanagi-visual_ex2}, only the peak pixel has a large value, which is considered to be
\add{a false detection caused by a noise spike,}
not the light source of the actual object.
\item 
For each of the detected objects, we calculate its coordinates on the CCD chip (referred to as the CCD coordinates).
Using the matching result with the Pan-STARRS1 PS1 $3\pi$ catalog described in Section \ref{sec:observation}, we convert the CCD coordinates into the equatorial coordinates.
\end{enumerate}

Tasks 7 and 8 are carried out on an FPGA board.
Typical analysis of 32 2K $\times$ 4K images with the 104 CCD chips of HSC (for the science use out of the total 116 chips) takes about 140 minutes.
In this study, we processed 14 datasets listed in Table \ref{tbl:obslog} with a total runtime of $140 \times 14 = 1960$ minutes, or $\sim 1.36$ day.

\onecolumn

\twocolumn

As previously mentioned, the algorithm described in this section was originally developed to detect space debris and near-Earth objects.
This algorithm requires a constant time interval between each image acquisition.
When we observe near-Earth objects, the typical exposure time of an image is relatively short, $\sim$ 20 seconds.
Therefore, a set of observations for several tens of images takes 20--30 minutes.
Thus, the use of a constant time interval throughout an entire set of observations is easily achieved.
On the other hand, the KBO observations at the Subaru Telescope may not continue image acquisition with a constant time interval because: refocusing is frequently performed, changes in weather conditions may interrupt the observations, and so on.
The number of images to be superimposed in our method is currently limited to 32 due to the restriction of the hard-coded algorithm embedded on FPGA \citep{yanagisawa2021}.
Since we cannot easily modify the code implemented on the FPGA, we prepare the 32 images with a constant time interval by deleting or duplicating some images when any irregular events occurred during the observation.

Table \ref{tbl:observation} shows the summary of our detections.
The number of moving object candidates detected from the whole dataset described in Table \ref{sec:observation} is 6980.
Among them, human vetting excluded 2641 candidates.
The remainder is $6980 - 2641 = 4339$ robust candidate detections of moving objects.
The detection limit through our method is 26.0--26.5 magnitudes (see Section \ref{sec:link}). 
The average detection limit of a single image with the exposure time of 90 seconds using HSC is 24--25 magnitudes \citep{aihara2018b}.
Therefore we can say that JAXA's superposition method is more capable of detecting fainter moving objects than ordinary detection methods using a single HSC image.
In Figure \ref{fig:apparentmagwhole} we show the frequency distribution of the apparent magnitude of the actual moving objects detected in this study.
We suspect that the faintest end in the distribution ($\sim$26.5 magnitude) indicates the detection limit of the dataset with our current method.

\onecolumn

\begin{figure}[!ht]\centering
  \includegraphics[width=0.84\textwidth]{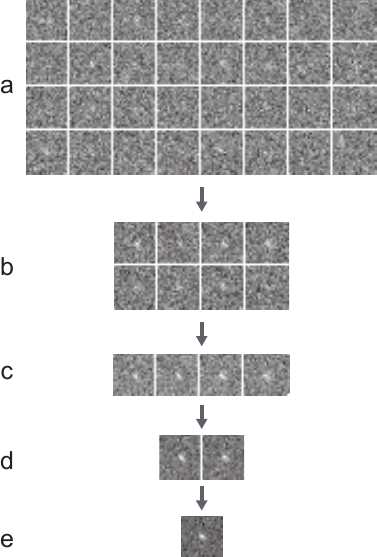}
\caption{%
An example set of visual inspection images that includes a moving object with little noise (the images are from the actual observation dataset).
The 32 square images in the panel \textsf{a} are part of the original 32 images that include the detected moving objects.
Panels \textsf{b}, \textsf{c}, \textsf{d}, and \textsf{e} represent the combination of 4, 8, 16, and 32 images with the median filter, respectively.
\label{fig:yanagi-visual_ex1}}
\end{figure}

\begin{figure}[!ht]\centering
  \includegraphics[width=0.84\textwidth]{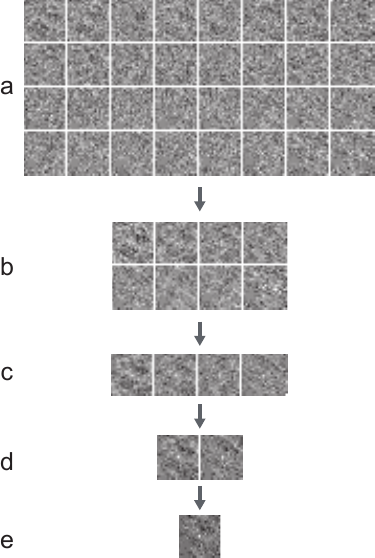}
\caption{%
An example set of visual inspection images of 
\add{a false detection caused by noise spike}
that does not include any moving objects (the images are from the actual observation dataset).
See the caption of Figure \ref{fig:yanagi-visual_ex1} for the meaning of each panel.
The light source at the center of the panel \textsf{e} is probably created by the overlap of high background noise.
\label{fig:yanagi-visual_ex2}}
\end{figure}

\twocolumn

\begin{table}\centering
\caption{%
Observation date (UT), Reduction ID, field, and the ratio between $N_\mathrm{real}$ (the number of real objects) and $N_\mathrm{detec}$ (the number of detections).}
\begin{tabular}{cccc}
Date (UT) & red. id & field & $N_\mathrm{real}/N_\mathrm{detec}$ \\
\hline
2020 May    28 & 03070 & F2 & $279/435$  \\ 
2020 May    29 & 03071 & F1 & $689/1042$ \\ 
2020 May    30 & 03072 & F2 & $727/918$  \\ 
2020 May    31 & 03073 & F1 & $316/404$  \\ 
2020 June   20 & 03093  & F2 & $54/198$   \\ 
2020 June   21 & 03094 & F1 & $77/207$   \\ 
2020 June   22 & 03095 & F2 & $286/457$  \\ 
2020 June   24 & 03097 & F1 & $151/227$  \\ 
2020 June   25 & 03098 & F2 & $316/459$  \\ 
2020 August 12 & 03146 & F1 & $251/354$  \\ 
2020 August 13 & 03147 & F2 & $448/712$  \\ 
2020 August 14 & 03148 & F1 & $495/1066$ \\ 
2021 June   09 & 03447 & F2 & $41/91$    \\ 
2021 June   17 & 03455 & F2 & $209/410$  \\ 
\hline
\end{tabular}
\label{tbl:observation}
\end{table}

\begin{figure}[ht!]
  \includegraphics[width=0.48\textwidth]{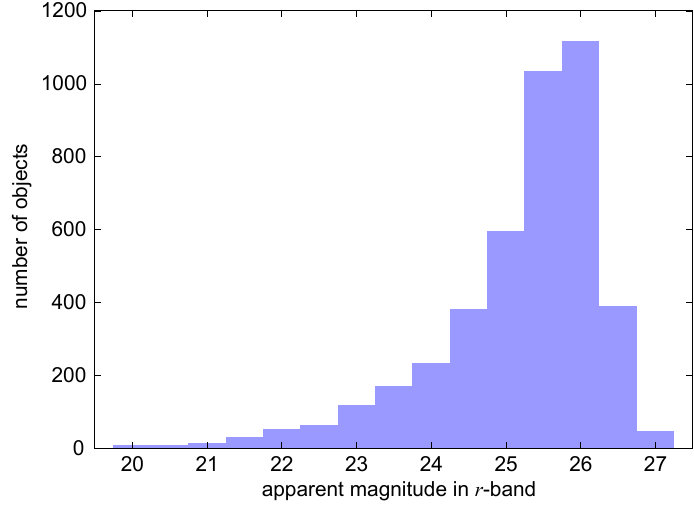}
\caption{%
Apparent magnitude distribution of the moving objects that we detected in this study over the entire observation period.
Note that we found five objects in the bin between 19.0 and 19.5 magnitude, which are outside of the plot range.
\label{fig:apparentmagwhole}}
\end{figure}

\section{Linking orbits} \label{sec:link}
In general, categorizing a moving object into one of the small body populations (such as KBOs) through their apparent velocity on the celestial sphere is possible only when the images are obtained near opposition \citep[e.g.][]{yoshida2003}.
Although our observational datasets range from May 2020 to June 2021, only the datasets acquired in June 2020 and June 2021 are along the direction of the opposition.
\add{%
We therefore searched for KBO candidates among the datasets obtained in June 2020 and June 2021.
In 2020, we also obtained datasets in May and August.
Thus, we examined whether the KBO candidates detected in the datasets obtained in June 2020 were also detected in the datasets obtained in May and August of that year.
We did it to improve the accuracy of our orbit determination by identifying KBO candidates over as long an observational arc as we could.
}  

\begin{figure}[ht!]
  \includegraphics[width=0.48\textwidth]{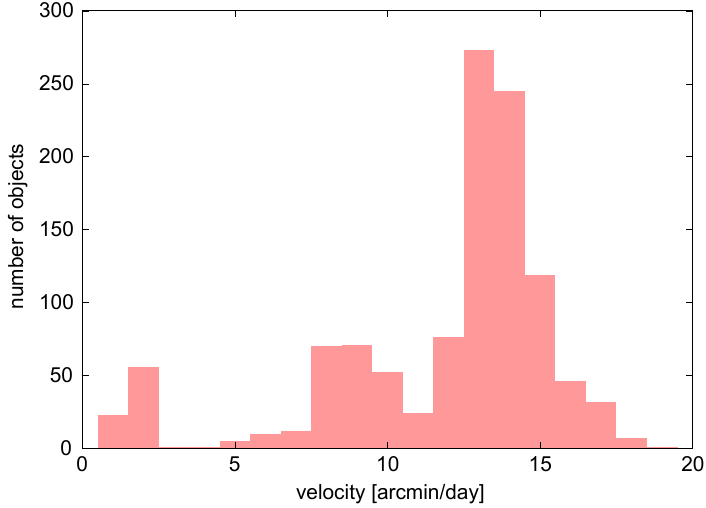}
\caption{%
Apparent velocity distribution of the moving objects detected in the observational datasets obtained in June 2020 and June 2021.
\label{fig:apparentvelocity}}
\end{figure}

\begin{figure}[ht!]
  \includegraphics[width=0.48\textwidth]{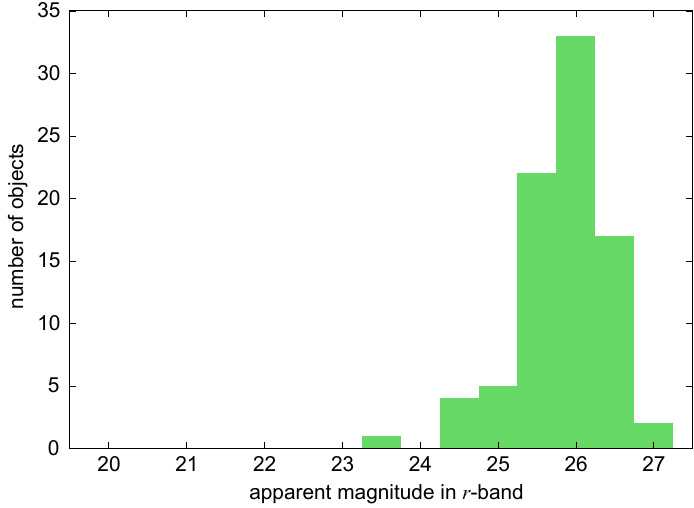}  
\caption{%
Apparent magnitude distribution of the KBO candidates detected in the June 2020 and June 2021 datasets.
\label{fig:apparentmagTNO}}
\end{figure}

In Figure \ref{fig:apparentvelocity} we show the apparent velocity distribution of moving objects detected in the datasets acquired in June 2020 and June 2021 when the survey areas were close to opposition.
The population of moving objects with a peak velocity $v$ distribution at $v \sim 13$ arcmin/day are main belt asteroids.
The objects having a velocity peak at 8--9 arcmin/day correspond to the L5 swarm of the Jupiter Trojans.
They coincidentally overlapped with our observation fields in June 2021.
The population with the peak velocity distribution of $v \lesssim 3.2$ arcmin/day \citep[or $v \lesssim 8$ arcsec/hour, e.g.][their Fig. 5]{yoshida2007} corresponds to KBOs.
In total, 84 KBO candidates were detected.
The observation in June 2020 was the closest to the opposition, so we extended the observation arc backward and forward using the June 2020 observation as the reference.
We have completed the identification of the KBO candidates for the observation period of May, June, and August 2020, and June 2021.
In Figure \ref{fig:apparentmagTNO} we show the apparent magnitude distribution of the KBO candidates detected in the June 2020 and June 2021 datasets when the survey areas were close to opposition. 

In what follows we describe how we link the orbits of objects detected over multiple nights.
Using the method described in Section \ref{sec:method}, we first calculate the apparent velocity $v$ of all the detected moving objects on the celestial sphere.
The velocities are determined from the values of the shift by the JAXA detection method.
We select KBO candidates from the detected moving objects using a velocity criterion of $v \lesssim 3.2$ arcmin/day.
For each KBO candidate, we calculate the values of its equatorial coordinates at the beginning and end of the single-night observation.
Using this information, we estimate their orbits and find objects whose orbits can be linked over multiple nights.
Here we use a software code called StellaHunter Professional.
\add{%
The StellaHunter Professional is the commercial PC software manufactured by a Japanese private company, AstroArts Co., Ltd (\url{https://www.astroarts.co.jp/}), mainly for amateur astronomers to detect moving objects in the solar system such as asteroids or comets from their observed images.
This software can determine the orbits of moving objects and find identical objects detected over multiple nights.
It has also implemented JAXA's moving object detection algorithm described in Section \ref{sec:method}, and realizes the detection of very faint objects.}
JAXA once provided the algorithm of the moving object detection to AstroArts, and AstroArts realized its implementation and commercialized the achievement as a software package (\url{https://www.astroarts.co.jp/products/stlhtp/index-j.shtml}, the webpage is written only in Japanese).
However, currently, this software is no longer available as a commercial product.
Readers who are interested in using it may want to contact the authors individually.

\begin{figure}[ht!]\centering
  \includegraphics[width=0.96\textwidth]{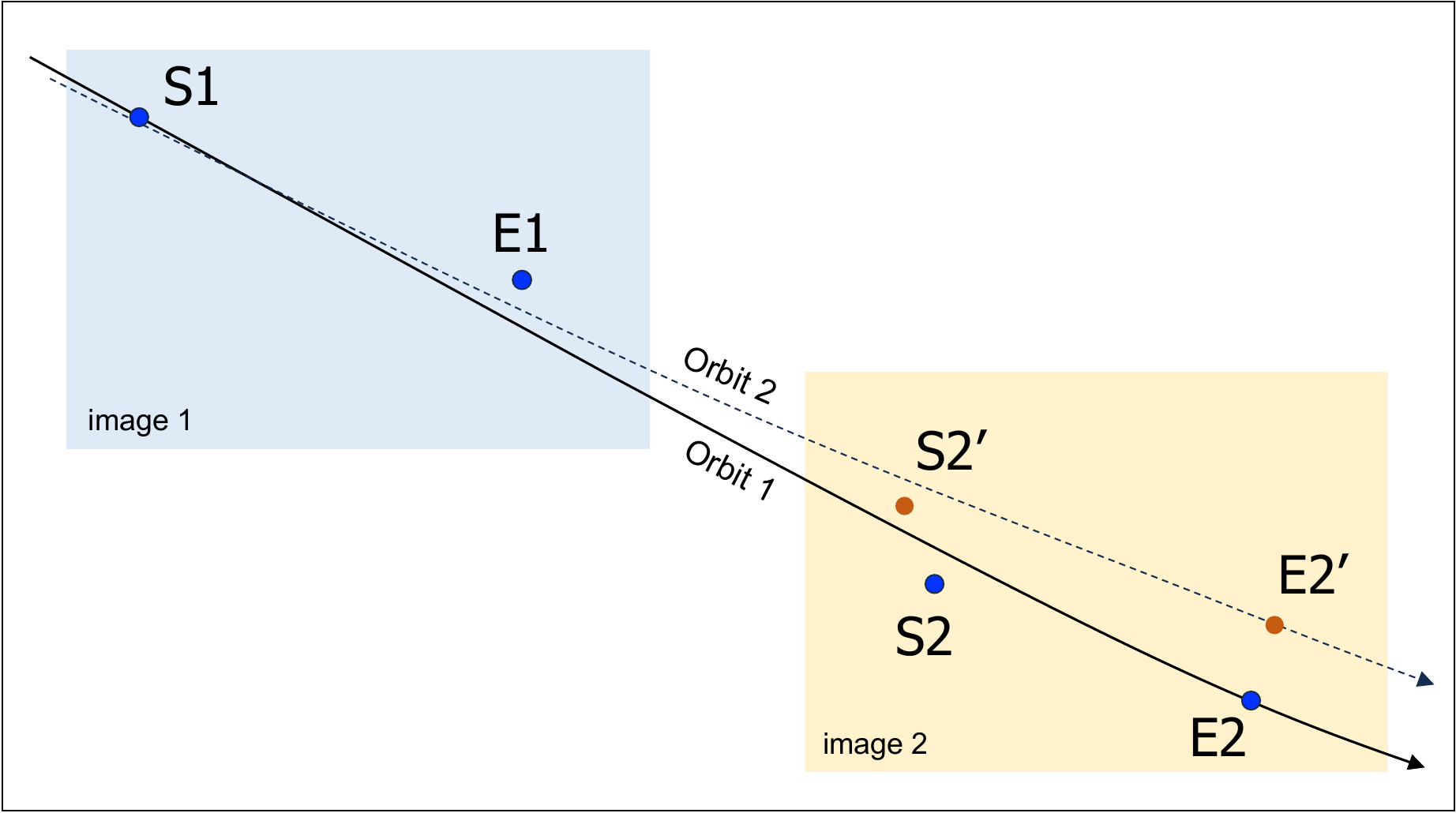}
\caption{%
Schematic illustration of how we link an object's orbit over multiple nights.
We draw just two images $(1, 2)$ for simplicity.
\textsf{S1} is the start position of an object candidate on the first night,
\textsf{E1} is its end position on the same night,
\textsf{S2} is the start position of one of the object candidates on the last night, and
\textsf{E2} is its end position on the same night.
On the other hand,
\textsf{S2$'$} is the start position of another object candidate on the last night, and
\textsf{E2$'$} is its end position on the same night.
\label{fig:schem-orbitlink}}
\end{figure}

We try many different multi-night combinations for each of the KBO candidates detected on each observation night and look for objects whose orbits could be successfully linked.
In what follows we itemize the tasks we conducted for this purpose.
See also Figure \ref{fig:schem-orbitlink} for a schematic illustration.
\begin{enumerate}
\item
We determine the Keplerian orbit that connects the start position (S1 in image 1 in Figure \ref{fig:schem-orbitlink}) and the end position (E2 in image 2 in Figure \ref{fig:schem-orbitlink}) of an object candidate over the combined nights.
This orbit is referred to as orbit 1 in Figure \ref{fig:schem-orbitlink}.
Position E2 may not belong to this object, but at this point, we presume it does.
\item 
We derive the deviation (root-mean square, the so-called O--C residual) of the positions E1 and S2 from orbit 1.
We call this residual 1.
Here we impose a condition that residual 1 must be smaller than twice the resolution of HSC (corresponding to the pixel scale of HSC), i.e. $< 0.34''$.
If residual 1 is larger than this value, we exclude the possibility that positions E2 and S2 belong to this particular object candidate and search for another candidate.
\item
If there are other object candidates in image 2, 
we determine the Keplerian orbit that connects the start position S1 and the end position of this candidate (E2$'$ in image 2 in Figure \ref{fig:schem-orbitlink}) over the combined nights.
This orbit is referred to as orbit 2.
\item 
We derive the O--C residual of the positions E1 and S2$'$ from Orbit 2.
We call it residual 2.
We impose the same condition for the value of residual 2 as that of residual 1, i.e. its value must be $< 0.34''$.
\item 
We see whether the residual 1 or residual 2 is larger.
\begin{itemize}
  \item If residual 1 is smaller than residual 2, we adopt orbit 1 as the orbit of the object candidate that belongs to the positions S1--E1 (and --S2--E2).
  \item If residual 1 is larger than  residual 2, we adopt orbit 2 as the orbit of the object candidate that belongs to the position S1--E1 (and --S2$'$--E2$'$).
\end{itemize}
\item When we carry out the orbit linking through the method we have described, 
sometimes the orbit of a KBO candidate (with low velocity) is linked to the orbit of an object of a very different population (such as a main belt asteroid).
To avoid this issue, in this study, we adopt another condition for successful orbit linking: the velocity and the magnitude of the two object candidates that make up a multi-night combination should be close to each other.
The specific thresholds that we use are as follows: the relative velocity difference of the two object candidates is less than $\pm 20${\%}, and the difference in their apparent magnitudes is less than 2 mag.
\end{enumerate}

As a result, we successfully linked six KBO orbits in 12 nights from May 2020 to August 2020.
In addition, we successfully linked the orbit of another KBO in two nights in June 2021.
\add{%
We estimate the orbital elements of these seven ($=6+1$) KBOs, and summarized their properties in Table \ref{tbl:detected}.
The meanings of the numbers in the Detection ID in Table \ref{tbl:detected} are as follows:
The first five digits (such as 03071) correspond to the Reduction ID in Table \ref{tbl:obslog}. The next three digits (such as 096) are the chip number of the CCD array of the HSC.
The last two digits (such as 02) are the sequential number of moving object candidates in order of its brightness on each CCD chip.
We numbered the moving object candidates detected on each CCD chip in order of brightness.
The brightest object is numbered 1, the second brightest is numbered 2, and so on.
The orbit of the object at the bottom with the Detection IDs 03447--027--01 and 03455--027--02 was linked in two nights in June 2021.
Here, 03447--027--01 denotes an object which is the brightest on the CCD chip 027 in the Reduction ID 03447 (which corresponds to the observation on 2021 June 9 as shown in Table \ref{tbl:observation}).
Similarly, 03455--027--02 denotes an object which is the second brightest on the CCD chip 027 in the Reduction ID 03455 (which corresponds to the observation on 2021 June 17).
}
Then we check whether they are newly discovered objects by using MPChecker \url{https://cgi.minorplanetcenter.net/cgi-bin/checkmp.cgi}.
We give MPChecker the following two conditions: search radius = $5'$, and limiting magnitude $V=30.0$.
MPChecker returns a report that the orbital elements of these seven objects are not close to any of the known objects.
Thus, it is likely that they are newly discovered.
\add{%
We report the detected position coordinates of these seven KBO candidates to the Minor Planet Center (MPC).}
\add{%
Amongst the seven objects, the MPC assigned provisional designations to two objects that had detections for more than three nights and observation arcs from May to August: 2020 KJ${}_{60}$ and 2020 KK${}_{60}$.}
We show the orbital elements calculated by the MPC for these two objects in Tables \ref{tbl:2020KJ60} and \ref{tbl:2020KK60}.
The following information is also appended to this table: 
first detection date ($t_\mathrm{detect}$),
Earth distance at $t_\mathrm{detect}$, and
Sun distance   at $t_\mathrm{detect}$.
We computed $t_\mathrm{detect}$ using JPL's Horizons System, \url{https://ssd.jpl.nasa.gov/}.

\onecolumn

\begin{landscape}
\begin{table}[htbp]\centering
\caption{%
Orbital properties of the KBOs detected in this study:
DATEOBS (UT),
Detection ID,
Epoch,
mean anomaly $M$ (deg),
mean motion $n$ (deg/day),
argument of perihelion $\omega$ (deg),
semimajor axis $a$ (au),
longitude of perihelion $\Omega$ (deg),
eccentricity $e$,
inclination $I$ (deg),
orbital period $P$ (year),
absolute magnitude $H$.
All the values were calculated in this study (not through MPC).
See the main text for the definition of Detection ID.
}
\begin{longtable}[c]{cccccccccccc}
\multicolumn{11}{c}{F1} \\
\hline
DATEOBS(UT) & Detection ID      & Epoch         & $M$       & $n$        & $\omega$  & $a$        & $\Omega$  & $e$       &  $I$     & $P$    & $H$   \\
\hline
2020/05/29 & 03071--096--02 &               &           &            &           &            &           &           &          &        &       \\
2020/06/21 & 03094--092--01 & 2020 May 31.0 &  11.28143 & 0.00220614 &  15.68541 & 58.4404403 & 242.23978 & 0.3964207 &  4.23433 & 446.77 &  8.53 \\
2020/08/12 & 03146--093--03 &               &           &            &           &            &           &           &          &        &       \\
\hline
2020/05/29 & 03071--080--02 & 2020 May 31.0 & 359.95345 & 0.00191703 &  90.65816 & 64.1775983 & 195.03697 & 0.3379206 &  2.68321 & 514.15 &  7.71 \\
2020/06/24 & 03097--074--01 &               &           &            &           &            &           &           &          &        &       \\
\hline
2020/05/29 & 03071--090--05 & 2020 May 31.0 & 179.82848 & 0.00706483 & 352.26576 & 26.8989165 & 113.28983 & 0.6167589 & 21.27055 & 139.51 &  9.61 \\
2020/06/24 & 03097--085--01 &               &           &            &           &            &           &           &          &        &       \\
\hline
2020/05/29 & 03071--094--11 & 2020 May 31.0 & 359.94102 & 0.00242844 & 124.95387 & 54.8173809 & 160.32114 & 0.2339671 & 3.52256  & 405.88 &  9.51 \\
2020/06/24 & 03097--089--01 &               &           &            &           &            &           &           &          &        &       \\
\hline
\multicolumn{11}{c}{F2} \\
\hline
2020/05/28 & 03070--066--06 &               &           &            &           &            &           &           &          &        &       \\
2020/05/30 & 03072--050--07 & 2020 May 31.0 & 359.77854 & 0.00298581 & 132.00232 & 47.7633370 & 154.91643 & 0.0822065 & 2.56450  & 330.11 &  9.39 \\
2020/08/13 & 03147--020--06 &               &           &            &           &            &           &           &          &        &       \\
\hline
2020/06/25 & 03098--014--04 & 2020 May 31.0 & 359.71567 & 0.00383246 &  20.64859 & 40.4406149 & 266.42684 & 0.0025594 & 5.15591  & 257.18 & 10.19 \\
2020/08/13 & 03147--014--06 &               &           &            &           &            &           &           &          &        &       \\
\hline
2021/06/09 & 03447--027--01 & 2021 July 5.0 &   0.03091 & 0.00174529 & 147.31747 & 68.3215293 & 139.02814 & 0.3493611 & 1.83469  & 564.75 &  8.96 \\
2021/06/17 & 03455--027--02 &               &           &            &           &            &           &           &          &        &       \\
\hline
\end{longtable}
\label{tbl:detected}
\end{table}
\end{landscape}

\twocolumn

\begin{table}[htbp]\centering
\caption{%
Properties of 2020 KJ$_{60}$.
The values are taken from the MPC website except for
first detection,
Earth Distance at $t_\mathrm{detect}$, and
Sun Distance at $t_\mathrm{detect}$ at the bottom.
$t_\mathrm{detect}$ and $t_\mathrm{detect}$ were calculated through JPL's Horizons System.
}
\begin{tabular}{lc}
\hline
epoch & 2023-09-13.0 \\
epoch JD & 2460200.5 \\
perihelion date	& 1984--01--18.13386 \\
perihelion JD & 2445717.634 \\
argument of perihelion (deg) & 69.3508 \\
ascending node (deg) & 156.60392 \\
inclination (deg) & 2.49898 \\
eccentricity & 0.3752396 \\
perihelion distance (au) & 37.2882832 \\
semimajor axis (au) & 59.6841373 \\
mean anomaly (deg) & 30.95781 \\
absolute magnitude & 9.67 \\
phase slope & 0.15 \\
uncertainty & 9 \\
observations used & 6 \\
oppositions & 1 \\
arc length (day) & 77 \\
residual rms (arcsec) & 0.08 \\
first observation date used & 2020--05--28.0 \\
last observation date used & 2020--08--13.0 \\
\hline
first detection ($t_\mathrm{detect}$, UTC) & 2020--05--28 11:18 \\
Earth Distance at $t_\mathrm{detect}$ & 42.542 au \\
Sun Distance   at $t_\mathrm{detect}$ & 43.328 au \\
\hline
\end{tabular}
\label{tbl:2020KJ60}
\end{table}

\begin{table}[htbp]\centering
\caption{%
Same as Table \ref{tbl:2020KJ60}, but about 2020 KK$_{60}$.
}
\begin{tabular}{lc}
\hline
epoch & 2023-09-13.0 \\
epoch JD & 2460200.5 \\
perihelion date	& 2006--03--16.60277 \\
perihelion JD & 2453811.103 \\
argument of perihelion (deg) & 15.16474 \\
ascending node (deg) & 242.43004 \\
inclination (deg) & 4.24758 \\
eccentricity & 0.3888881 \\
perihelion distance (au) & 35.261981 \\
semimajor axis (au) & 57.7013481 \\
mean anomaly (deg) & 14.36763 \\
absolute magnitude & 8.77 \\
phase slope & 0.15 \\
uncertainty & 9 \\
observations used & 6 \\
oppositions & 1 \\
arc length (day) & 75 \\
residual rms (arcsec)	& 0.16 \\
first observation date used & 2020--05--29.0 \\
last observation date used & 2020--08--12.0 \\
\hline
first detection ($t_\mathrm{detect}$, UTC) & 2020--05--29 10:44 \\
Earth Distance at $t_\mathrm{detect}$ & 35.667 au \\
Sun Distance   at $t_\mathrm{detect}$ & 36.474 au \\
\hline
\end{tabular}
\label{tbl:2020KK60}
\end{table}

\section{Discussion} \label{sec:discussion}

In this study, we analyzed the observational dataset originally acquired for the New Horizons target search at the Subaru Telescope using JAXA's moving object detection method.
Our analysis covers the datasets obtained between May 2020 and June 2021.
However, the categorization of a moving object into a unique solar system population using solely apparent velocity is only possible when observations are executed near opposition.
Therefore we paid particular attention to the datasets acquired in June 2020 and in June 2021 when the observation field was close to the opposition.
Then we found 84 KBO candidates.

The series of observations for New Horizons/KEM
started in May 2020, and is on-going.
In this study, we only analyzed the datasets obtained between May 2020 and June 2021 because they have past the proprietary period and are publicly available (note that the data proprietary period for the Subaru Telescope is 18 months).
More datasets acquired in this series of observations will be released to the public in due course as the proprietary period expires. 
We will continue to analyze additional data as they become available in the archive.
The primary goal of this effort is, of course, to find the next in-situ observational candidate for the New Horizons spacecraft.
Yet it simultaneously leads to a more detailed understanding of the dynamical structure of the deepest region of the solar system, particularly the orbital distribution of the small bodies dominated by diverse resonances \citep[e.g.][]{saillenfest2019a,ito2019,malhotra2019}
\add{and the distribution of cometary objects in the inner part of the Oort Cloud \citep[e.g.][]{dones2015,fouchard2018,fouchard2023}.}

One of the unique perspectives of observing KBOs from a spacecraft flying through the Kuiper Belt is that
they can be observed with a significantly larger solar phase angle, and from a much closer distance, compared with ground-based or Earth-based observation.
For example, the New Horizons spacecraft observed the large classical KBO
(50000) Quaoar at solar phase angles of $51^\circ$, $66^\circ$, $84^\circ$ and $94^\circ$ \citep{verbiscer2022}.
Ground-based observation can only provide data at small solar phase angles ($\lesssim2^\circ$).
The combination of observations at large and small solar phase angles provides us with knowledge of the surface reflectance of the object and enables us to infer information about a KBO's surface properties in detail \citep[e.g.][]{porter2016,verbiscer2018,verbiscer2019,verbiscer2022}.

As we continue our work, we are aware that several issues need to be addressed.
First, the number of KBOs detected with our method as reported this time may not be optimal, and it could have been even larger using the same dataset.
More specifically, we suspect the mask patterns that we used in this study can be improved in the future.
As explained in Section \ref{sec:method}, we subtract the median image from each image created across all the images.
This process should remove all the field stars, ideally leaving only moving objects. 
However, in reality, the effect of incomplete subtraction remains due to the difference in the seeing, mainly caused by atmospheric turbulence.
To deal with this problem, we applied the mask patterns to remove the effect of incomplete subtraction.
The values of the masked regions are set to 0, which means the regions are useless for detecting moving objects.
Since the observed sky regions (F1 and F2) are crowded with field stars, the regions of the mask patterns occupy almost 20{\%} of the entire images.
This could eliminate the detection of KBOs that exist near those field stars.
In the near future, although it will require significant computational resources, we plan to use the subtracted images and consider seeing differences in each image, and apply them to our detection method. 
Partly for this purpose, we plan to execute our detection method on a GPU cluster, which we expect will reduce the turn-around time of the data analysis to $\sim 1/5$ or less compared with the present
(note that this is based on a test analysis obtained using Nvidia Quadro RTX8000 $\times 4$).

The second issue is related to the efficiency of our object detection procedure.
As discussed in Section \ref{sec:method}, our detection method still relies heavily on visual inspection of the events at the end of the procedures (``human vetting'', the task no. 10 in the flowchart in Figure \ref{fig:yanagi-flowchart}).
This becomes a formidable task as the amount of data increases, and it is obvious that sooner or later the workload from this task will exceed the limit of human capability.
Recently, use of machine learning has become commonplace in the field of astronomy, in particular for detection of moving objects in the solar system \citep[e.g.][]{lin2018,lee2022,jia2023}.
We have begun preparation to exploit the advantage of machine learning and will bring a machine-learning framework into JAXA's detection method in the near future as a substitute for human vetting.

The third issue is about the number of superimposed images that JAXA's detection method uses.
Currently, the method uses 32 images as a set (see Section \ref{sec:method}).  This number can be 
increased to 64 by improving our software.
The result would be the ability to reach a much fainter (deeper) level; we are preparing for this improvement now.

\section{Acknowledgments}
This research is based on data collected at the Subaru Telescope operated by the National Astronomical Observatory of Japan (NAOJ), and the data is obtained from SMOKA operated by Astronomy Data Center (ADC), NAOJ.
We are honored and grateful for the opportunity of observing the Universe from Maunakea, which has the cultural, historical, and natural significance in Hawaii.
The observations were supported in part by NASA for the New Horizons mission.
In particular, NASA Keck exchange time was used to acquire the data on 17 June 2021 (S21A--TE216--K).
F.~Y. acknowledges support from the Japan Society for the Promotion of Science (JSPS) Kakenhi grant, 20H04617.
J.~J. acknowledges support from the Japan Society for the Promotion of Science (JSPS) Kakenhi grants, JP19K23456 and JP22K14069.
A.~V. acknowledges support from JPL/RSA contract no. 1659537.
StellaHunter Professional is a trademark of AstroArts.
This study has made use of NASA's Astrophysics Data System (ADS) Bibliographic Services.

\bibliography{mybib}{}
 \bibliographystyle{plainnat-pasj} 

\end{document}